\documentclass[a4paper,12pt]{article}
\newcommand{\be}{\begin{equation}}
\newcommand{\ee}{\end{equation}}
\newcommand{\bea}{\begin{eqnarray}}
\newcommand{\eea}{\end{eqnarray}}
\newcommand{\bean}{\begin{eqnarray*}}
\newcommand{\eean}{\end{eqnarray*}}

\newcommand{\ra}{\rangle}
\newcommand{\lra}{\leftrightarrow}
\newcommand{\bc}{\begin{center}}
\newcommand{\ec}{\end{center}}
\newcommand{\btab}{\begin{tabular}}
\newcommand{\etab}{\end{tabular}}
\newcommand{\hl}{\hline}
\newcommand{\nn}{\nonumber}
\def\qq{$ q\bar q $}
\def\ss{$ s\bar s $}
\def\nn{$n\bar{n}$}
\def\uu{$u\bar u$}
\def\dd{$d\bar d$}
\begin{document}
\begin{titlepage}

\baselineskip=18pt \vskip 0.9in
\begin{center}
{\bf \Large Radiative decays: a new flavour filter}\\
\vspace*{0.3in}
{\large F.E. Close}\footnote{\tt{e-mail: f.close@physics.ox.ac.uk}} \\
\vspace{.1in}
{\it Department of Theoretical Physics,
University of Oxford, \\
Keble Rd., Oxford, OX1 3NP, United Kingdom}\\
\vspace{0.1in}
{\large A. Donnachie}\footnote{\tt{e-mail: ad@a35.ph.man.ac.uk}} \\
{\it Department of Physics and Astronomy, University of Manchester}\\
{\it Manchester M13 9PL, United Kingdom}\\
\vspace{0.1in}
{\large Yu.S. Kalashnikova}\footnote{\tt{e-mail: yulia@heron.itep.ru}} \\
{\it ITEP}\\
{\it Moscow, Russia}\\
\end{center}

\begin{abstract}
\noindent Radiative decays of the $1^3D_1$ orbital excitations of the $\rho$, 
$\omega$ and $\phi$ to the scalars $f_0(1370)$, $f_0(1500)$ and $f_0(1710)$ 
are shown to provide a flavour filter, clarifying the extent of glueball
mixing in the scalar states. A complementary approach to the latter is 
provided by the radiative decays of the scalar mesons to the ground-state 
vectors $\rho$, $\omega$ and $\phi$. Discrimination among different mixing 
scenarios is strong.
\end{abstract}
\end{titlepage}

\subsection*{Introduction}

The observation and confirmation of gluonic degrees of freedom in mesonic
states is of great significance for QCD, which predicts the existence of
glueballs (bound states of gluons) and of hybrids (quark-antiquark-gluon
states). Calculations in lattice QCD give estimates of the likely masses of
glueballs \cite{glue} and light-quark hybrids \cite{hybrids}, and evidence
for the excitation of gluonic degrees of freedom has emerged in several
processes. Lattice calculations predict that the lightest glueball has
$J^{PC}=0^{++}$ and is in the mass range 1.3 to 1.7 GeV. Experimentally
\cite{PDG} there is one more $0^{++}$ state in this mass range than can be
accomodated by excited quark-antiquark states. The natural inference is that
there is a glueball state present \cite{Am95,CK02a}. A clear exotic resonance,
$\pi_1(1600)$, with quantum numbers $J^{PC}=1^{-+}$ which cannot be accessed
by a pure quark-antiquark state, has been seen \cite{E852a} in the $\eta'(958)\pi$
channel in the reaction $\pi^-N \to (\eta'(958)\pi)N$ and in the $\rho^0\pi^-$
channel \cite{E852b,VES} in the reaction $\pi^-N \to (\pi^+\pi^-\pi^-)N$.
Anomalously large hadronic decay modes of light-quark vector mesons, that is
decays which are predicted to be very small in standard models \cite{BCPS},
are observed in $e^+e^-$ annihilation around 1.6 GeV. A favoured explanation
\cite{DK93,CP97,DK99} is to include vector hybrids which, on the basis of the
$\pi_1(1600)$ mass, are expected to be in this mass region.

Apart from states with exotic quantum numbers, disentangling hybrids and
glueballs from quark-antiquark states is difficult using only hadronic
decay channels because of mixing. However, radiative transitions offer
special opportunities. The coupling to the charges and spins of constituents
reveals detailed information about wave functions and discriminates among
models. In the case of gluonic excitations of the $\pi$ and $\rho$, that is
hybrids, the spin structure differs from conventional excitations with the
same overall $J^{PC}$. For example, in a hybrid $1^{--}$ the \qq~ are in a
spin-singlet, while for the $0^{-+}$ they are in a spin-triplet: in each
case this is the reverse of what one is accustomed to. In the case of the
scalar mesons
a direct measure of their electromagnetic couplings gives
information about the flavour content of 
the scalar states and could resolve
the issue of $G$-$q\bar q$ mixing.

We have calculated \cite{CDK02} the rates for the radiative decays $V^* \to
\gamma M$, where the $V^*$ are the $2S$ and $1D$ excitations of the $\rho$,
$\omega$ and $\phi$ in the 1.4 to 1.8 GeV mass range and the $M$ are positive
C-parity $q\bar{q}$ states in the 1.2 to 1.7 GeV mass range. We label the 
vector states $\rho_S$, $\rho_D$, $\omega_S$, $\omega_D$, $\phi_S$ and 
$\phi_D$. Some of these vector decays are predicted to have branching ratios 
in excess of $10^{-2}$ and will be measurable at existing and planned 
facilities, such as $e^+e^-$ annihilation by Initial State Radiation (ISR) 
at Babar and Belle, by direct $e^+e^-$ annihilation at the upgraded VEPP 
collider at Novosibirsk and by diffractive photoproduction following the 
upgrade at Jefferson Laboratory.

These radiative transitions serve two purposes, exploring the nature both
of the initial excited vector state and that of the resultant meson. In
\cite{CDK02} we concentrated on the former aspect. Here we emphasise the
latter and extend the calculation to the radiative decays of the scalars
to the ground-state vector mesons. There is again great potential for precise
measurement of these decays, for example at CLEO via the decay $J/\psi \to
\gamma\gamma V$ or at COMPASS via central production in high-energy proton
proton collisions.

Radiative transitions of $\rho^*$ or $\omega^*$ to $\gamma M$ couple directly
to the \uu$\pm$\dd~ content of $M$. In similar vein, the analogous transitions
involving $\phi^*$ couple to the \ss~ content of $M$. By comparing the
relative rates for isovector ($|M\ra \equiv$ ${{1}\over{\sqrt{2}}}
(|$\uu$\ra$ - $|$\dd$\ra$) and isoscalar ($|M\ra \equiv \cos \theta$
$|$\ss$\ra + \sin \theta$ ${{1}\over{\sqrt{2}}}(|$\uu$\ra$ + $|$\dd$\ra$))
for a set of mesons $M$ of positive C-parity and the same $J^P$ it is
possible in principle to determine the relative amount of \ss~ and non-strange
flavours in the $M$ wavefunction and hence to weigh the flavour content of the
nonet. In the case where a glueball has mixed into the multiplet extending it
to a decuplet, as is hypothesised \cite{Am95,CK02a} to be the case for the 
$f_0(1370)$, $f_0(1500)$ and $f_0(1710)$, such transitions enable the
role of the glueball to be disentangled and might even be able to estabish the
mass of the ``quenched'' glueball \cite{CK02a}. Furthermore, as we shall show,
radiative decays of the scalar mesons to the ground-state vectors $\rho(770)$
and $\phi(1020)$ are also sensitive probes of the glueball mixing in the
scalars.

\subsection*{The model}

The details of the radiative decay calculation are described in \cite{CDK02}.
Wavefunctions for the excited vector mesons are found variationally from the
Hamiltonian
\be
H = \frac{p^2}{m_q} + \sigma r - \frac{4}{3} \frac{\alpha_s}{r} + C
\label{hamilton}
\ee
with standard quark model parameters $m_q = 0.33$ GeV for $u$ and $d$ quarks
and 0.45 GeV for $s$ quarks, $\sigma = 0.18$ GeV$^2$ and $\alpha_s = 0.5$.
The wavefunctions are taken to be Gaussian of form $\exp(-p^2/(2\beta_M^2))$
multiplied by the appropriate polynomials and $\beta_M$ treated as the
variational parameter in $H$ for each of the $1S,1P,2S,1D$ states. Where data
already exist good agreement was found.

For transitions involving excited states with nodes in their wavefunctions
there is {\it a priori} concern about the stability of the calculations. This
does not arise in the transitions we consider here, and the approximation of
harmonic oscillator wave functions is `safe'. This can be seen, for example,
in the $1P \to 1S$ transitions. Using the operator relation $\vec p=im_q[H 
\vec r]/2$, where $H$ is the Hamiltonian (\ref{hamilton}), one obtains
for the radial integrals of the wave functions the equality
\be
\langle B | p | A \rangle = \frac{im_q}{2}\omega\langle B | r | A \rangle,
\ee
where $A$ is the initial meson, $B$ is the final meson, $\omega=m_A-m_B$ 
and $m_A$ and $m_B$ are the model, not the physical masses.
With Gaussian wave functions the radial integrals (up to a phase) are
\be
\langle B | p | A \rangle = {{1}\over{\sqrt{2}}}{{\beta^5}\over
{\beta_A^{5/2}\beta_B^{3/2}}}
\ee
and
\be
\langle B | r | A \rangle = 
{{1}\over{\sqrt{2}}}{{\beta_A^{5/2}\beta_B^{3/2}}
\over{\beta_0^5}}
\ee
where 
\be
\beta^{-2}={{1}\over{2}}(\beta_A^{-2}+\beta_B^{-2})
\label{betadef}
\ee
and
\be
\beta_0^2={{1}\over{2}}(\beta_A^2+\beta_B^2) 
\label{beta0def}
\ee
From \cite{CDK02},
$\beta_A=0.274$ GeV, $\beta_B=0.313$ GeV and the mass difference
$\omega = 0.562$ GeV. So we find
\be
{{\langle A | p | B \rangle}\over{m_q}}=0.658
\label{pav1}
\ee
and
\be
{{\omega\langle A | r | B \rangle}\over{2}}=0.616
\label{rav1}
\ee
The deviation between (\ref{pav1}) and (\ref{rav1}) is less than 10
percent so, as the squares enter the widths, the uncertainty due to
the use of Gaussian wave functions is about 20 percent. This is
appreciably less than the effects we are considering.

A similar exercise for $1D \to 1P$ gives
\be
{{\langle B | p | A \rangle}\over{m_q}}=1.088
\label{pav2}
\ee
and
\be
{{\omega\langle B | r | A \rangle}\over{2}}=1.024
\label{rav2}
\ee
so in this case the uncertainty due to the use of Gaussian wave
functions is less than 10 percent.

The pure electric-dipole ($E1$) transition is well-defined for heavy
quarks but is certainly a bad approximation for light quarks. We
include the magnetic quadrupole ($M2$) transition as well and there is
a long history of success with this approach, even though the $M2$ terms
are at the same order in $p^2$ as $E1$ corrections proportional to
the anomalous magnetic moment of the constituents, spin-orbit
terms, Thomas precession and binding effects\cite{leyou,lit}. 
Some of these corrections can be calculated \cite{leyou,lit,calc,gk,gko}; 
some can only be estimated \cite{leyou}. In any case, going beyond the 
leading approximation for the electric and magnetic amplitudes requires 
knowledge of the Lorentz nature of the confinement force.

The most notable successes of this approach, namely of calculating the 
leading terms for each of the relevant electric and magnetic multipoles, 
have been in reproducing the magnitudes and relative phases of over 100 
helicity amplitudes for photoexcitation of the proton and neutron 
\cite{baryon}. These give a clear indication of which amplitudes are 
large or small, and of their relative sizes and signs. This success 
suggests that although corrections may be individually significant, 
their collective effect is small. 

It is possible to go beyond such an approximation, but at the price of 
losing some predictive power. Within the general assumption that 
electromagnetic amplitudes are additive in the constituents, it is 
possible to obtain relations among the helicity amplitudes, angular 
distributions and widths for a set of states by normalising the reduced
amplitude to some subset of observables \cite{fecbook,melosh}. 

This is the philosophy that we shall adopt. First, within the ``leading 
multipole" hypothesis we can make two checks of our procedures. 

\noindent (i) We find that our result for the decay $f_1(1285) \to \gamma
\rho$ is in good accord with experiment \cite{PDG,MARK3,WA102}. 

\noindent (ii) We predict that
$\Gamma(f_2(1270) \to \gamma\rho) < \Gamma(f_1(1285) \to \gamma\rho)$,
which appears also to be in
accord with experiment in that there is no evidence for the radiative decay of
the $f_2(1270)$ in either the MARKIII \cite{MARK3} or WA102 \cite{WA102,AK}
experiments, and both have strong $f_2$ signals.

Secondly, within the ``single-quark-transition" hypothesis we can form a 
positivity constraint among a combination of the widths. This is satisfied 
by our explicit model dependent calculations, but enables us to draw a more 
general conclusion, namely that
$\Gamma(f_0 \to \gamma \rho) \sim \Gamma(f_1 \to \gamma \rho)$.

\subsection*{Vector meson decays}

Our interest here is in the radiative decays of the $\rho_D$, which we identify
with the $\rho(1700)$ \cite{PDG}, to $f_0(1370)$ and $f_0(1510)$ and of the
(unobserved) $\phi_D$, to which we assign a mass of about 1.9 GeV, to
$f_0(1710)$. In the absence of glueball mixing we predict \cite{CDK02}
$\Gamma(\rho_D \to \gamma f^{n\bar{n}}_0(1370)) \sim 900$ keV,
$\Gamma(\rho_D \to \gamma f^{n\bar{n}}_0(1500)) \sim 600$ keV 
(assuming that one or the other is a pure $n\bar{n}$ state) and
$\Gamma(\phi_D \to \gamma f^{s\bar{s}}_0(1710)) \sim 200$ keV. These
widths can be changed substantially when glueball mixing is included, the
degree of modification depending on the mass of the bare glueball.

Three different mixing scenarios have been proposed: the bare glueball is
lighter than the bare $n\bar n$ state \cite{CK01}; its mass lies between
the bare $n\bar n$ state and the bare $s\bar s$ state \cite{CK01}; or it
is heavier than the bare $s\bar s$ state \cite{LW00}. We denote these three
possibilities by L, M and H respectively. In \cite{CDK02} we identified
potentially powerful ways of determining glueball mixing in the scalar mesons
through the radiative transitions $\rho_D \to \gamma f_0(1370)$ and $\gamma
f_0(1500)$. If the bare glueball is light ($\sim 1300$ MeV) it will mix
strongly with the $f^{n\bar{n}}_0(1370)$ and dilute the 900 keV width which
will be pushed into the other scalars, in particular into the
$f^{n\bar{n}}_0(1500)$ reversing the relative magnitudes of the radiative widths.
At the other extreme, the mixing of a heavy glueball ($\sim 1700$ MeV) does
not materially affect the radiative width to the $f_0(1370)$ but severely
depresses that to the $f_0(1500)$ to 100 keV or less. Thus the relative
widths of $\rho_D \to \gamma f_0(1370)$ and $\rho_D \to \gamma f_0(1500)$
are sensitive to the glueball mass. There is similar sensitivity in the decays
$\phi_D \to \gamma f_0(1500)$ and $\phi_D \to \gamma f_0(1710)$. For a light
glueball the former is essentially zero and the latter is predicted to be
$\sim 170$ keV. For a heavy glueball the situation is reversed, with the decay
$\phi_D \to \gamma f_0(1500)$ predicted to be about 260 keV and the
decay $\phi_D \to \gamma f_0(1710)$ to be essentially zero.
These branching ratios are challenging but important as they open the
possibility of weighing the \nn~ and \ss~ flavour content of the scalar
states and determining the bare mass of the scalar glueball. The only
question is whether experiment will be sensitive to such magnitudes.

In the isovector-scalar sector only the $a_0(980)$ is well-established.
The mass, and even the existence, of the $a_0(1450)$ remains controversial.
Its mass is a critical parameter in mass matrices for the mixing of the
scalar glueball with the \qq~ nonet. We predict $\Gamma(\omega_D \to \gamma
a_0(1450)) \sim 610$ keV and $\Gamma(\rho_D \to \gamma a_0(1450)) \sim 85$ keV.
Of course in $e^+e^-$ annihilation or diffractive photoproduction the
$\rho_D$ will be produced at approximately nine times the rate of the
$\omega_D$, so the $\gamma a_0(1450)$ rates will be the same for both.
Observation of the $a_0(1450)$ in radiative decays would have significant
implications, particularly when coupled with the radiative decays to the
isoscalar scalars.

Although our emphasis is on the radiative decays to scalars as a probe
of their glueball content it is important to look at other decays as a
check on the model. We find that the principal radiative decay mode of 
the $\rho_D$ is $\gamma f_1(1285)$, with an estimated width \cite{CDK02} 
of $\sim 1100$ keV. Although the decays of the $f_1(1285)$ are many-bodied, 
principally $4\pi$ and $\eta\pi\pi$, this is compensated for by its narrow 
width of 24 MeV. The largest radiative decay of the $\omega_D$ is found to 
be $\gamma a_1(1260)$, with a width $\sim 1100$ keV. Observation of these 
modes would confirm our calculations. The E852 experiment at Brookhaven has 
observed \cite{bnl1640} the decay $\omega(1640) \to \omega\eta$. If the 
$\omega(1640) \equiv \omega_D$ then, in the $^3P_0$ model, the partial 
width for this mode is predicted \cite{BCPS} to be 13 MeV. Thus, if the 
$^3P_0$ model is a good guide in this case, a measureable $\omega(1640) 
\to \gamma a_1$ is anticipated: as E852 have several thousand events in 
the $\omega \eta$ channel, it is possible that there could be several 
hundred $\gamma a_1$ events in their experiment.

\subsection*{Scalar meson decays}

Just as the radiative decays of excited vector states to the scalars provide
sensitive tests of their parton content, so do the radiative decays of the
scalars to the ground-state vectors. In our model these decays are given by
\be
\Gamma(f_0 \to \gamma V)=\frac{8}{3} \alpha p \frac{E_B}{m_A} \frac{\beta^2}{m^2_q}F^2
(1+\lambda \frac{p^2}{\beta^2})^2I
\label{f0}
\ee
where $A$ refers to the initial scalar meson, $B$ to the final vector ($V$) 
meson,
\be
\lambda={{\beta_A^2}\over{2(\beta_A^2+\beta_B^2)}}
\ee
\be
F={{\beta^4}\over{\beta_A^{5/2}\beta_B^{3/2}}}\exp\Big(-{{p^2}\over
{8(\beta_A^2+\beta_B^2)}}\Big)
\ee
where $\beta$ is given in (\ref{betadef}) and $I$ is an isospin factor, 
${{1}\over{4}}$ for $n\bar{n} \to \gamma\rho$
and ${{1}\over{9}}$ for $s\bar{s} \to \gamma\phi$. 

From \cite{CDK02}, $\beta_A = 0.274$ GeV, $\beta_B=0.313$ GeV for $n\bar n$
states and $\beta_A = 0.307$ GeV, $\beta_B=0.355$ GeV for $s\bar s$ states.
In the absence of glueball mixing it gives $\Gamma(f_0(1370) \to 
\gamma\rho) \sim 2300$ keV, and $\Gamma(f_0(1710) \to \gamma\phi) \sim 870$ 
keV. 

\vskip 0.1in

The width of the decay $f_1(1285) \to \gamma\rho$ is measured and provides
a check on the model. This decay is given by
\be
\Gamma(f_1 \to \gamma V)=\frac{8}{3} \alpha p \frac{E_B}{m_A} \frac{\beta^2}{m^2_q}F^2
(1+\lambda \frac{p^2}{\beta^2}+\frac{1}{2}\lambda^2 \frac{p^4}{\beta^4})I
\label{f1}
\ee
which
for an $n\bar{n}$ state
 predicts a width of 1400 keV. This compares well with the experimental
value \cite{PDG,MARK3,WA102} of $1320 \pm 312$ keV. 

\vskip 0.1in

The decay $f_2(1270) \to \gamma\rho$
is given by
\be
\Gamma(f_2 \to \gamma V)=\frac{8}{3} \alpha p \frac{E_B}{m_A} \frac{\beta^2}{m^2_q}F^2
(1-\lambda \frac{p^2}{\beta^2}+\frac{7}{10}\lambda^2 \frac{p^4}{\beta^4})I
\label{f2}
\ee
which is rather smaller than the others, due to the negative 
contribution of the $p^2/\beta^2$ term. This predicts a width of 644 keV.

Experimentally this width is small as
neither the MARKIII \cite{MARK3} nor the WA102 \cite{WA102} experiments
has any evidence for it. The branching fractions for the radiative decay
of  $J/\psi$ to $f_1(1285)$ and $f_2(1270)$ are comparable \cite{PDG}, at
$(6.1 \pm 0.9)\times 10^{-4}$ and $(1.39 \pm 0.14)\times 10^{-3}$
respectively, so the non-observation of any $f_2(1270)$ signal in the
decay $J/\psi \to \gamma(\gamma\rho)$ is meaningful. A similar
situation holds in central production in high-energy proton-proton
interactions\cite{WA102}
 and one can deduce \cite{AK} an upper limit on $\Gamma(f_2
\to \gamma\rho)$ of 500 keV at 95\% confidence level. 
So it is reasonable to suppose that our results
for the $f_0$ radiative decays are valid.

As in \cite{CDK02} we consider three scenarios for the glueball mass:
light, medium and heavy. The predicted widths for the decays of
$f_0(1370)$, $f_0(1500)$ and $f_0(1710)$ to $\gamma\rho$ and $\gamma\phi$
are given in table 1.

\bigskip

\bc
\btab{|c|ccc|ccc|}
\hl
&&$\rho(770)$&&&$\phi(1020)$&\\
&L&M&H&L&M&H\\
\hl
$f_0(1370)$&443&1121&1540&8&9&32\\
\hl
$f_0(1500)$&2519&1458&476&9&60&454\\
\hl
$f_0(1710)$&42&94&705&800&718&78\\
\hl
\etab
\ec
Table 1. Effect of mixing in the scalar sector of the $1^3P_0$ nonet
for radiative decays to $\rho$ and $\phi$. The radiative widths,
in keV, are given for three different mixing scenarios as described
in the text: light glueball (L), medium-weight glueball (M) and heavy
glueball (H). 
\bigskip

It is clear from the table that the discrimination among the different
mixing scenarios is strong. The decay $f_0(1500) \to \gamma\rho$ is
perhaps the most interesting because of its comparatively narrow total
width of $\sim 120$ MeV. This enhances the radiative decay branching
fraction, which is $\sim 2\%$ for a light glueball, $\sim 1\%$ for a
medium glueball and $\sim 0.5\%$ for a heavy glueball.
The absolute magnitudes may vary by $\sim 20\%$ due to the uncertainties in
the model, but the relative strengths manifested in the pattern above should 
be robust. Even allowing for the intrinsic uncertainties, there appears to be 
a solid conclusion that these rates are worth pursuing experimentally.
 
We now assess further the robustness of these results by looking at the
more general structures that follow from the single-quark-transition 
property of the dynamics.

\subsection*{Single quark transition}

The results above depend upon the rather general assumption that 
electromagnetic amplitudes are additive in the constituents. Independent 
of details of binding dynamics, it is possible to obtain
relations among the helicity amplitudes, and hence angular distributions and 
widths, that depend only on the assumption that the mesons are described as 
$q\bar{q}$ states in the $P$ and $S$ states respectively. 

For example, in the electric-dipole approximation, the relative widths for 
transitions from $f_J \to \gamma \rho$ would be equal, apart from phase space 
corrections. Given the magnitude of one as input, the others immediately 
follow. While such results are a good approximation for heavy flavours 
\cite{PDG}, for the lighter $u,d,s$ flavours, significant magnetic-quadrupole
may be expected, as is confirmed by the explicit calculations above.
(Electric octupole contributions, while allowed in general for $f_2 \to 
\gamma V$, will vanish if, as assumed here, the vector $V$ is 
a pure $^3S_1$ state, and the $f_2$ is a pure $^3P_2$ state \cite{fecbook}.)

Within
this assumption that the electromagnetic transition amplitude is additive
in the constituents \cite{fecbook,melosh} and that the relevant $q\bar{q}$ 
states can be classified as $S,P$ levels,
sum rules can be obtained. We can of course check that these sum rules 
are satisfied 
in particular explicit models.

The most general decomposition of the helicity amplitudes for ${^3}S_1 
\lra {^3}P_J$ radiative transitions into electric-dipole and 
magnetic-quadrupole terms is given in \cite{gk,gko}. The helicity amplitudes
may be written as: 

${^3}S_1 \lra {^3}P_0$
\be
A_0 = \sqrt{2}(E_1+2E_R)
\ee
\vskip 2pt
${^3}S_1 \lra {^3}P_1$
\be
A_0 = \sqrt{3}(E_1+E_R+M)~~A_1=\sqrt{3}(E_1+E_R-M)
\ee
\vskip 2pt
${^3}S_1 \lra {^3}P_2$
\be
A_0 = (E_1-E_R+3M)~~A_1=\sqrt{3}(E_1-E_R+M)~~A_2=\sqrt{6}(E_1-E_R-M)
\ee

Here $E_1$ is the leading electric-dipole term, $E_R$ is the ``extra''
electric-dipole term and M is the magnetic-quadrupole term. In terms of
our model, $E_1$ comes from the convection current and $E_R$ and
$M$ from the spin-flip one (see equation (2) of \cite{CDK02}). The 
width for the decays $V \to \gamma f_J$ is
\be
\Gamma(V \to \gamma f_J) = {{p^3}\over{3}}\sum_\lambda |A_\lambda|^2
\ee
and for the decays $f_J \to \gamma V$ is
\be
\Gamma(f_J \to \gamma V) = {{p^3}\over{2J+1}}\sum_\lambda |A_\lambda|^2 
\ee
Consider first the decays $V \to \gamma f_J$. For equal phase space and 
equal form factors, we can eliminate the $|M|^2$ term by the combination 
$3\Gamma(V \to \gamma f_1) - \Gamma(V \to \gamma f_2)$. This leaves cross 
terms between $E_1$ and $E_R$ which are in general of indeterminate sign. 
However, these too can be eliminated by including $\Gamma(V \to \gamma f_0)$ 
and a combination formed that is proportional to the sum $|E_1|^2 + |E_R|^2 
\geq 0$.

Thus, for equal phase space and equal form-factors, one would have
\be
\Gamma(\rho(2S) \to \gamma f_2) + 7\Gamma(\rho(2S) \to \gamma f_0) \geq
3\Gamma(\rho(2S) \to \gamma f_1) 
\label{radial}
\ee

Similarly for the decays of pure $n\bar{n}$ $f_J$ states,
\be
5\Gamma(f_2 \to \gamma \rho) + 7\Gamma(f_0 \to \gamma \rho) \geq
9\Gamma(f_1 \to \gamma \rho) 
\label{ground}
\ee

It is straightforward to verify that (\ref{radial}) is satisfied by
equations (17) to (19)\footnote{There is a sign error in equation (18) in 
\cite{CDK02}. The sign of the $G_E G_M$ term should be negative, as it is 
in equation (A5).} of \cite{CDK02} and that (\ref{ground}) is satisfied 
by (\ref{f0}), (\ref{f1}) and (\ref{f2}) of the present paper. 

Given that empirically $\Gamma(f_1 \to \gamma \rho) \sim 1500$ keV, equation 
(\ref{ground}) implies that one or other of the $f_2$ or $f_0$ must have a 
radiative width of at least $\sim 1000$ keV. To the extent that there is no 
clear sign of the  $f_2 \to \gamma \rho$ decay in either the data sets of 
\cite{MARK3,WA102} one may anticipate that the $f_0 \to \gamma \rho$ could 
be large, in line with our specific calculations. In any event, such states 
should be sought in  $\psi \to \gamma \gamma \rho$.
 
\subsection*{Conclusions}

Our computations show that radiative decay rates can be large for some
transitions. The limitation for detection of these processes seems to be
one primarily of acceptance. These processes should be borne in mind when
detectors for future experiments, for example at Hall D in Jefferson Lab,
are being designed. We advocate that existing data be mined to seek evidence
of radiative decays, for example in E852 where the decay $\omega(1640) \to
\gamma a_1$ should be present at a reasonable rate. If radiative transitions
are observed, then the use of radiative decays will be proven as a viable
technique.

It is important to compare photoproduction and $e^+e^-$ annihilation
as these give complementary information on the vector-meson wavefunctions.
Jefferson Laboratory is an obvious place for the photoproduction studies,
and as regards $e^+e^-$, VEPP is the immediate natural candidate, though
there is uncertainty as to whether its luminosity will be enough. DAFNE
operating up to its maximum energy promises high luminosity in the future.
Initial state radiation at BaBar, Belle and CLEO-c could give significant
rates for $e^+e^-$ annihilation at $\sqrt{s} \leq 2$ GeV.

A complementary approach to the role of glue in the scalar wave functions
is provided by the decays of the scalars to $\gamma\rho$ and $\gamma\phi$.
Both the $f_0(1500)$ and $f_0(1710)$ have been clearly observed in the
radiative decays $J/\psi \to \gamma X$, and we recommend that a detailed 
exploration of the scalar states should be planned as part of the CLEO-c 
programme.

Radiative transitions provide a new flavour filter that can clarify the
nature of the scalar mesons in the 1.2-1.8 GeV region, in particular the
role of glue in the scalar wavefunctions.

\newpage
\bc
{\bf Acknowledgements}
\ec

This work is supported, in part, by grants from the Particle Physics and
Astronomy Research Council, RFBR 00-15-96786, INTAS-RFBR 97-232 and the
EU-TMR program ``Eurodafne'', NCT98-0169.

\end{document}